# Web-Based Question Answering: A Decision-Making Perspective

**David Azari**
University of Washington
Seattle, Washington
azari@cs.washington.edu

**Eric Horvitz**
Microsoft Research
Redmond, Washington
horvitz@microsoft.com

**Susan Dumais**
Microsoft Research
Redmond, Washington
sdumais@microsoft.com

**Eric Brill**
Microsoft Research
Redmond, Washington
brill@microsoft.com

**Abstract**

We investigate the use of probabilistic models and cost-benefit analyses to guide the operation of a Web-based question-answering system. We first provide an overview of research on question-answering systems. Then, we present details about AskMSR, a prototype question-answering system that synthesizes answers from the results of queries to a Web search engine. We describe Bayesian analyses of the quality of answers generated by the system and show how we can endow the system with the ability to make decisions about the nature and number of queries that should be issued, by considering the expected value and cost of submitting the queries. Finally, we review the results of a set of experiments.

## 1 Introduction

For several decades, researchers have pursued the goal of developing computational machinery with the ability to generate answers to freely-posed questions. General question-answering systems depend on techniques for analyzing questions and for composing answers from some corpus of knowledge. This is a challenging problem because the corpus may not contain an explicit matching answer or may contain multiple variants of relevant answers or answer components.

We have been interested in procedures that enlist the poorly-structured but copious resources of the Web for answering questions. Web-based question answering systems typically employ rewriting procedures for converting components of questions into sets of queries posed to search engines, along with techniques for converting query results into one or more answers.

To date, there has been little understanding of the value of alternate query rewriting strategies and answer composition methods. We have also had little knowledge about the enhancement of the quality of answers with the issuance of increasing numbers of queries to search engines. Given the burden that widely fielded question-answering systems can place on search engines, gaining a deeper understanding of the nature and number of query rewrites is important for deploying real-world question-answering systems.

We describe an investigation of probabilistic modeling and decision analyses to characterize and control querying policies in Web-based question answering. We first provide a brief review of prior work on question answering. We focus particularly on a system developed at Microsoft Research, named AskMSR. We present the rewrite procedures and answer composition methods performed by AskMSR. Then, we describe a set of experiments we undertook to better understand how alternative rewrite methods influenced the ultimate quality of answers. Beyond exploring alternate procedures, we also studied the influence of the quantity of rewrites on the quality of answers. Such an analysis relies on an effective strategy for ordering queries by their expected value, so as to allow the learning of models that can reason about the costs and benefits of employing additional numbers of queries to the web. We describe the methods we developed and review a set of experiments that demonstrate the effectiveness of the cost-benefit procedures.

## 2 Question-Answering Systems

Most text retrieval systems operate at the level of entire documents. In searching the web, complete web pages or documents are returned. There has been a recent surge of interest in finer-grained analyses focused on methods for obtaining *answers* to *questions* rather than retrieving potentially relevant documents or best-matching passages from queries—tasks information retrieval (IR) systems typically perform. The problem of question answering hinges on applying several key concepts from information retrieval, information extraction, machine learning, and natural language processing (NLP).

Automatic question answering from a single, constrained corpus is extremely challenging. Consider the difficulty of gleaning an answer to the question *"Who killed Abraham Lincoln?"* from a source which contains only the text *"John Wilkes Booth altered history with a bullet. He will forever be known as the man who ended Abraham Lincoln's life."* As Brill et al. (2002) have shown, however, question answering is far easier when the vast resources of the Web are brought to bear, since hundreds of Web pages contain the literal string *"killed Abraham Lincoln."*

### 2.1 Approaches to Question Answering

The TREC Question Answering Track (*e.g.,* Voorhees & Harman, 2000) has motivated much of the recent work in the field of question answering. The initial efforts in question answering have focused on fact-based, short-answer questions such as *"Who killed Abraham Lincoln?"*, *"What was the length of the Wright brothers first flight?"*, *"When did CNN begin broadcasting"* or *"What two US biochemists won the Nobel Prize in medicine in 1992?"*

Question-answering systems have typically used NLP analyses to augment standard information retrieval techniques. Systems often identify candidate passages using IR techniques, and then perform more detailed linguistic analyses of both the question and matching passages to find specific answers. A variety of linguistic resources (part-of-speech tagging, parsing, named entity extraction, semantic relations, dictionaries, WordNet, etc.) are used to support question answering. The Falcon system by Harabagiu et al. (2001) is typical of the linguistic approaches and has demonstrated excellent performance in benchmark tests. In the system, a query is parsed to identify important entities and to suggest a likely answer type. A rich taxonomy of answer types has been developed using lexico-semantic resources from WordNet (Miller, 1995). WordNet represents more than 100,000 English nouns, verbs, adjectives and adverbs into conceptual synonym sets, as encoded by lexicographers over the course of many years. Candidate matching paragraphs are similarly analyzed to see if they match the expected answer type. Often, relevant passages will not share words with the query. In these cases, the Falcon system uses WordNet to examine morphological alternatives, lexical alternatives (*e.g.,* nouns "killer," "assassin," or "slayer" will match the verb "killed"), and semantic alternatives (*e.g.*, "cause the death of"). Additional abductive processes are also used to provide answer justification and rule out erroneous answers.

### 2.2 Web Question Answering

In contrast to these rich natural language approaches, others have developed question answering system that attempt to solve the difficult matching and extraction problems by leveraging large amounts of data. AskMSR (Brill et al., 2002; Dumais et al., 2002) is an example of such a system, and one that we explore in more detail in this paper. The main idea behind the AskMSR system is to exploit the redundancy provided by the web to support question answering. Redundancy, as captured by multiple, differently phrased answer occurrences, facilitates question answering in two important ways. First, the larger the information source, the more likely it is that answers bearing close resemblance to the query can be found. It is quite straightforward to identify the answer to *"Who killed Abraham Lincoln?"* given the text, *"John Wilkes Booth killed Abraham Lincoln in Ford's theater."* Second, even when no exact answer can be found, redundancy can facilitate the recognition of answers by enabling procedures to accumulate evidence across multiple matching passages.

Other researchers have also looked to the web as a resource for question answering. The Mulder system (Kwok et al., 2001) is similar to AskMSR in many respects. For each question, Mulder submits multiple queries to a web search engine and analyzes the results. Mulder does sophisticated parsing of the query and the full-text of retrieved pages to identify answer candidates. Mulder also employs a local database of term weights for answer extraction and selection. Mulder has not been evaluated with TREC queries, so its performance is difficult to compare with other systems.

Clarke et al. (2001) investigated the importance of redundancy in their question answering system. They found that the best weighting of passages for question answering involves using both passage frequency (what they call redundancy) and a global term weight. They also found that analyzing more top-ranked passages was helpful in some cases and not in others. Their system builds a full-content index of a document collection, in this case the TREC test collection. Their implementation requires an auxiliary web corpus be available for full-text analysis and global term weighting.

Kwok et al. (2001) and Clarke et al. (2001) perform complex parsing and entity extraction for both queries and best matching web pages, which limits the number of web pages that they can analyze in detail. They also require term weighting for selecting or ranking the best-matching passages which requires auxiliary data structures. AskMSR is distinguished from these in its simplicity and efficiency. The system only uses simple rewrites and string matching, and makes direct use of summaries and simple ranking returned from queries to web resources. The data-driven techniques perform well in TREC benchmark tests (Voorhees and Harman, 2001).

## 3 AskMSR Prototype

We now turn to reviewing details of the operation of AskMSR as background for our efforts to extend the heuristic system via introducing probabilistic models and automated decision making. After reviewing AskMSR, we will describe our work to develop Bayesian models of the performance of AskMSR components, and to integrate cost-benefit strategies for guiding the system's actions.

The design of AskMSR was motivated by several efforts within NLP research that have demonstrated that, for many applications, significant improvements in accuracy can be attained by significantly increasing the amount of data used for learning. Following the same guiding principle, the tremendous data resources that the Web provides was used as the backbone of AskMSR.

AskMSR contains two main components, *query rewriting* and *answer composition*, which consists of several sub-processes (see Brill et al., 2002; Dumais et al., 2002 for details).

### 3.2 Query Rewriting

AskMSR reformulates each user question into likely substrings of declarative answers to the question. For each question, several rewrites are generated using eight rewrite heuristics. The rewrites vary from specific string matching to a simple "ANDing" of all the query words. As an example, for the query *"Who killed Abraham Lincoln?"* there are three rewrites: <LEFT> "killed Abraham Lincoln"; "Abraham Lincoln was killed by" <RIGHT>; and who AND killed AND Abraham AND Lincoln. <LEFT> and <RIGHT> refer to the likely placement of candidate answers. The first two rewrites require that a text on the Web match the exact phrase, such as "killed Abraham Lincoln." We refer to the last rewrite as a conjunctional *back-off strategy*, as it simply "ANDs" together all the query words, leading to less specific queries.

The rewrite strings are formulated as search engine queries and sent to a search engine from which page summaries are collected. Any search engine can be used as the provider of results to the second stage of AskMSR's analysis. AskMSR assigns heuristic scores to results of different kinds of rewrites. The system assigns higher weights to the results of more precise rewrites than it does to the more general back-off rewrite.

### 3.3 Answer Composition

Several phases of analysis are employed in AskMSR to identify answers to questions from the results returned by searches with query rewrites.

*Mine N-Grams.* From the page summaries returned for each query rewrite, all unigram, bigram and trigram word sequences are extracted. The *n*-grams are scored according to their frequency of occurrence and the weight of the query rewrite that retrieved it. As an example, the common *n*-grams for the example query about the assassination of Abraham Lincoln are: Booth, Wilkes, Wilkes Booth, John Wilkes Booth, bullet, actor, president, Ford's, Gettysburg Address, derringer, assignation, etc.

*Filter N-Grams.* The *n*-grams are filtered and re-weighted according to how well each candidate matches the expected answer type, as specified by fifteen handwritten filters. These filters use surface-level string features, such as capitalization or the presence of digits. For example, for *When* or *How many* questions, answer strings with numbers are given higher weight, and for *Who* questions, answer strings with capitals are given added weight and those with dates are demoted.

*Tile N-Grams.* Finally, the *n*-grams are *tiled* together by lining up matching sub-phrases where appropriate, so that longer answers can be assembled from shorter ones. Following tiling, the answers to the example query are: John Wilkes Booth, bullet, president, actor, Ford. John Wilkes Booth receives a much higher score than the other answer candidates because it is found in matches to specific rewrites and because it occurs more often overall.

## 4 Challenge of Limiting Query Costs

AskMSR's performance has been judged in the TREC question-answering conference to be competitive with the best question answering systems (Voorhees and Harman, 2001). The system can be viewed as deriving its power by employing relatively simple strategies targeted at leveraging the redundancy of the informational content of the Web. Unfortunately, the same mechanisms which provide its power make the system's operation costly. AskMSR generates an average of 7 rewrites per query (in the test collections to be described below). Large scale deployment of the system to many simultaneous users would place a significant burden on the backend search engines.

We set out to explore the possibility of using machine learning to better understand the value of different kinds of rewrites, and to build models that could be used to control the classes and numbers of query rewrites issued to search engines. This work involves understanding how the probability of identifying a correct answer is influenced by the properties of the question and the nature and number of queries issued to search engines.

In addition to providing guidance on the policies for generating and submitting queries, models of accuracy and cost could enable the system to know when it would be best to skip completing the pursuit of an answer to a question. In these cases, the system could instead ask a user to attempt a reformulation of the question, or to seek the answer elsewhere.

Finally, beyond seeking characterization and control, probabilistic analyses of accuracy and value of alternate query policies could also lead to new insights with implications for refining the methods used by the base system.

Brill et al. (2002) explored a related problem of using learning techniques to estimate the confidence the system has in the answer. However, Brill et al. did not explore the quality of individual rewrites, the quantity of rewrites allowed, or perform a cost-benefit analysis as we have.

## 5 Analysis of Answer Quality

Research on the analysis and control of the heuristic processes of the AskMSR system is facilitated by the system's architecture. AskMSR processes a question in distinct stages in a question-answering pipeline that can be independently analyzed. We set out to learn about the query reformulation and $n$-gram mining stages of the pipeline, with an eye on controlling the nature and numbers of queries issued to search engines.

### 5.1 Understanding the Value of Queries

In the pursuit of limiting the number of queries issued by AskMSR, we sought to replace the expert-derived heuristic functions used in AskMSR with Bayesian models that could generate probabilities of success.

In an initial phase of analysis, we explored models that could provide a ranking of individual queries. Our work on developing scores of query value was stimulated by our direct inspection of query rewrites generated by the system; many of the rewrites appeared to be nonsensical. We sought to endow AskMSR with insight about poor queries.

We employed Bayesian learning procedures to generate models from a training set of cases that could be used to infer the probabilistic lift in accuracy that queries of different types would confer. Such models promised to provide a normalized metric for ordering sets of queries by their value, providing a decision surface for deliberating about the costs and benefits in a more global analysis of the end-to-end performance of the overall AskMSR system.

### 5.2 Establishing a Query-Quality Gradient

We first separated queries into two categories: (1) queries that involve ANDing of individual words and occasionally short phrases (e.g., population AND "of Japan"), and (2) queries that contain a single phrase (e.g., "the population of Japan is"). We refer to the former as *conjunctional rewrites*. We refer to the latter as *phrasal rewrites*. These two sets of queries have several distinct features, which we examined in our modeling efforts.

For both types of rewrites, we considered such features as the number of distinct words and the number and percentage of stop words present in the queries. For building predictive models of the goodness of *phrasal rewrites* we additionally examined similar features, but also included features derived from a statistical natural language parser for English text created by the Natural Language Group at Microsoft. The syntactic parser constructs multiple parse trees, capturing multiple hypotheses for an input string, based on a consideration of the likely different parts of speech that words in a phrase can have. After producing all hypotheses, the parser employs a language model to rank the likely syntactic hypothesis, computing probabilities of each parse tree as the product of the probability of all of the nodes in the tree.

The application of NLP parsing to each query rewrite does not put a significant computational burden on clients hosting AskMSR. Rewrites are parsed on an order of milliseconds.

We took into consideration several features output by the parser including the number of primary and secondary parses and the maximum probability parse tree, or a measure of grammatical "goodness" of a query rewrite. A complete list of the features used for both sets of query rewrites is listed in Tables 1 and 2.

Table 1: Features of conjunctional and phrasal rewrites considered in learning models of query goodness.

| |
|---|
| LONGPHRASE: The longest phrase in the rewrite, in terms of words. |
| LONGWD: The length of the longest word in the entire query. |
| NUMCAP: The number of capitalized words in the entire query. |
| NUMPHRASES: The total number of phrases in the overall query. |
| NUMSTOP: The number of stopwords in the entire query, using our list. |
| NUMWORDS: The number of words in the entire query string. |
| PCTSTOP: Percentage of stop words. |

Table 2: Features used only for phrasal rewrites considered in learning models.

| |
|---|
| NUMCAP, NUMSTOP, PCTSTOP: as above. |
| PRIMARY_PARSES: The number of primary parses given by the natural language parser. |
| SECONDARY_PARSES: The number of secondary parses given by the natural language parser. |
| SGM: The "statistical goodness" of the rewrite; a measure of how grammatical the sentence or phrase is, given by the parser. |

We employed the WinMine toolkit for Bayesian learning developed by Microsoft Research (Chickering et al., 1997) to train decision models for the query rewrites from a training set. To generate training cases, we ran AskMSR on questions included in the TREC-9 data set. This data set includes a set of questions and correct answers used in the annual TREC workshop for evaluating the performance of competing question-answering systems (Voorhees & Harman, 2000). For each query, we collected rewrites generated by AskMSR for the TREC-9 data set. Cases were created by examining the features of conjunctional and phrasal query rewrites provided by the system (as shown in Tables 1 and 2), and noting the success of the system in answering the questions with single queries. The accuracy of the models was tested with questions drawn from TREC-10 questions.

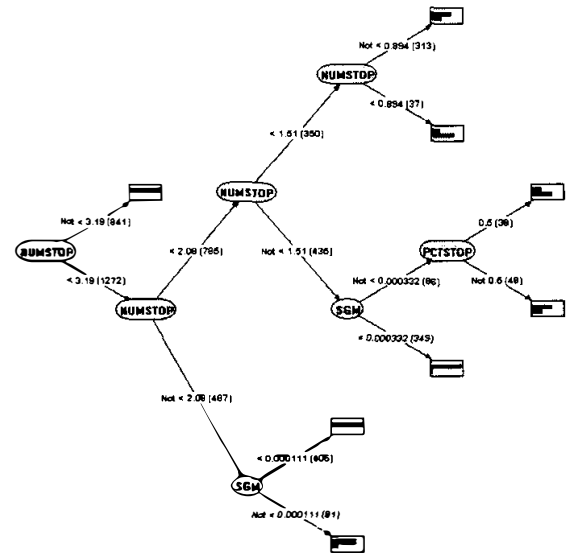

Figure 2: Decision tree for predicting success of phrasal queries.

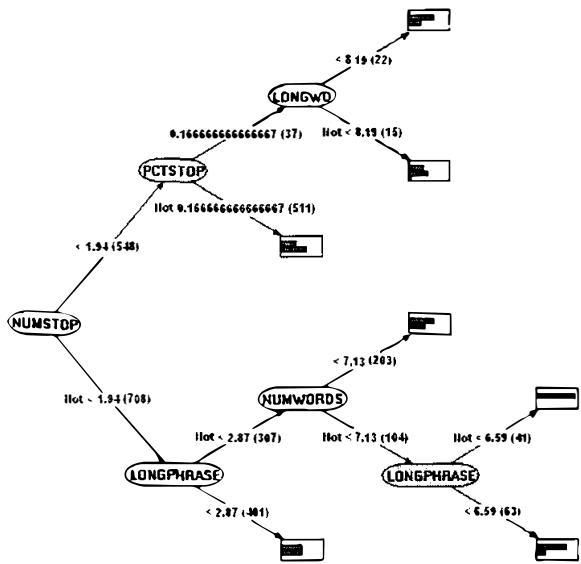

Figure 1: Decision tree learned for predicting success of conjunctional queries.

Figure 1 displays a decision tree derived from the Bayesian model that maps properties of queries based on conjunctional rewrites to an expected accuracy of answers to questions when a conjunctional query is issued to the Web. Figure 2 depicts a model for the accuracy of questions associated with the issuance of queries derived from phrasal rewrites. These models provide the probabilities that specific single rewrites will lead to a correct answer to a question. We use the probabilities that individual queries will achieve a correct answer as a *query-quality* score for ordering the list of rewrites in a subsequent analysis. The ordering provides a decision surface for a cost-benefit analysis of the ideal number of queries to issue. The ordering is heuristic in that the system does not use single queries in normal operation, but rather ensembles of queries.

### 5.3 Learning the Influence of Quantity on Quality

The initial analysis, yielding models of the usefulness of individual rewrites, enabled us to build a new version of AskMSR that orders the submission of queries according to the probability that individual queries will provide an accurate answer. We set out, in a second stage of learning and analysis, to understand how best to control to control the numbers of queries issued by the revised version of AskMSR.

In the second phase of analyses, we again use machine learning to build Bayesian models of the relationship between the ultimate accuracy of AskMSR's processing of questions and the numbers of queries submitted to a search engine, considering the properties of the question at hand. Such models enable cost-benefit analyses, trading off the expected gains in accuracy of an answer with the costs of submitting additional queries. These analyses provide AskMSR with new abilities for making dynamic decisions about the number of queries to submit to a search service—and to make decisions about when to forego an analysis and, instead, to ask a user to reformulate their question. We built an ensemble of models by generating cases via a process of running AskMSR on TREC questions and applying different fixed thresholds on the number of rewrites submitted to search engines, as ordered by the goodness of queries established in the first phase of model construction. Additional features used in this phase are shown in Table 3.

We note that the threshold numbers of rewrites were not always submitted because some questions generated fewer rewrites than the threshold values allowed.

Table 3: Features considered by the models for choosing rewrite thresholds for a given question-answering run.

| |
|---|
| AVERAGE_SNIPPETS_PER_REWRITE: Snippets are the summaries collected from web pages for a given query. |
| DIFF_SCORES_1_2: The difference between the first and second highest scored answer from AskMSR's scoring heuristic. |
| FILTER: The filter applied to the original query, such as "nlpwin_who_filter". |
| FILTER2 : Filters that focus on words and bigrams. |
| MAXRULE: Scores are given at the reformulation stage, based on the filter used to generate rewrites. This is the highest score procured for a particular query. |
| NUMNGRAMS: Total ngrams mined from snippets. |
| RULESCORE_X: Number of ngrams for rules with score X. |
| STD_DEVIATION_ANSWER_SCORES: The std. deviation amongst the top five answer scores from AskMSR's heuristic. |
| TOTALQUERIES: Total queries issued after all rewrites. |
| TOTNONBAGSNIPS: Total snippets generated from *phrasal* rewrites. |
| TOTSNIPS: Total snippets for all rewrites. |

In our experiments, we discretized the number of queries into fixed thresholds at 1-10, 12, 15, and 20 rewrites per question, thus building 13 models. The models generated by this process provide predictions about the overall accuracy of answers to questions at increasingly higher levels of thresholds on query rewrites submitted to a back-end search engine. Figure 3 displays a decision tree learned from data about the performance of question answering when limiting submitted queries to 10 rewrites.

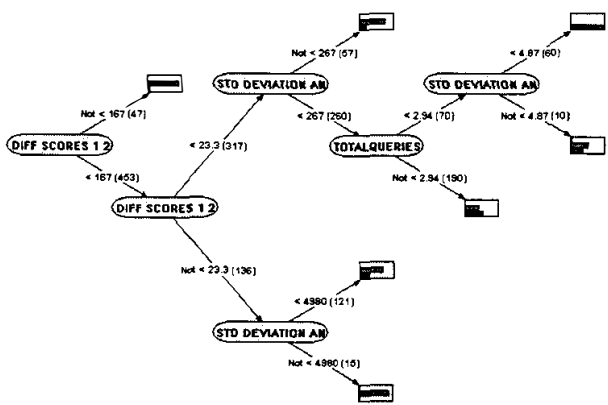

Figure 3: Decision tree for a query rewrite threshold of 10 rewrites per question. Models such as this were constructed for 1-10, 12, 15, and 20 maximum rewrites per question.

## 6 Cost-Benefit Considerations

Once we generate a set of Bayesian models that can predict the ultimate accuracy of answers to questions for different numbers of query rewrites, we are poised to deploy a system with the ability to dynamically control the number of queries used to answer previously unseen questions. We refer to the new version of AskMSR as AskMSR-DT (for AskMSR-Decision Theoretic).

In controlling the number of queries relayed to a search engine, we need to represent preferences about the costs of sending increasing numbers of queries to search engines and the benefits of obtaining a more accurate answer. Several models for representing costs and benefits are feasible.

We considered a model where a user or system designer assesses a parameter $v$, indicating the dollar value of receiving a correct answer to a question, and a parameter $c$ representing the cost of each query rewrite submitted to a search engine. Rather than asserting a constant value for receiving an answer to a question, a user may consider the value of receiving an answer as a function of the details of the situation at hand. For example, the value of an answer may be linked to the type of question, goals, and even the time of day for a user. Likewise the cost of submitting queries can a function of such factors as the as the current load sensed on a search engine or the numbers of queries being submitted by a user's entire organization to a third-party search service. Costs may also be asserted directly as a fee for query by a search service. The costs may be linear in the number of queries or may scale non-linearly with increasing numbers of queries. For example, the first $n$ queries may be considered free by a search service supporting the question-answering systems at an enterprise, after which expenses are incurred in a supra-linear manner.

Models that output the probability of retrieving a successful answer, conditioned on different numbers of query rewrites, allow us to compute the expected value of submitting the queries. If we take the value of not receiving an valid answer as zero, the expected value of submitting $n$ queries is the product of the likelihood of the answer, given evidence $E$ about the query and background state of knowledge $\xi$, $p(A|E,n,\xi)$, and the value of obtaining a correct answer $v$, $p(A|E,n,\xi)\,v$.

Let us take the simple example of a preference model where the value of an answer, $v$, is assessed in terms of the cost of queries, $c$. That is, we assess the value of answers as some multiple $k$ of the cost of each query $c$, $v=kc$.

In a deployed version of AskMSR-DT, a user or system administrator for an enterprise could be provided with an easy-to-use interface for assessing preferences about value and costs. Easy access to such controls would allow users to change preferences about the willingness to pay for accuracy in different settings.

Let us assume a cost model that grows linearly with the number of queries, $nc$. In making decisions about the ideal number of queries to submit, we seek to optimize the net expected value, computed as the difference of the expected value and cost, for different $n$. Thus we wish to find the ideal number of queries, $n^*$ where

$$n^* = \arg \max_n p(A|E,n,\xi)kc - nc.$$

AskMSR-DT has the ability to check each quantity of query rewrites explored in the machine learning studies, and identify the best number of queries to submit.

Figure 4 shows an idealized view of the case of a cost-benefit analysis where the probability of an answer grows with decreasing marginal returns with additional query reformulations. The expected value, cost, and net expected value are displayed as a function of the number of queries submitted. If we had such smooth decreasing marginal returns on accuracy with increasing numbers of queries, we could identify $n^*$ simply from the derivatives of the curves. As indicated in Figure 4, the ideal number of queries to issue is obtained at a position on the $x$-axis where the change in expected value of the answer is equal to the cost of each query. In reality, given the potential non-monotonicity of the expected value curve, we check the number of queries associated with each learned model.

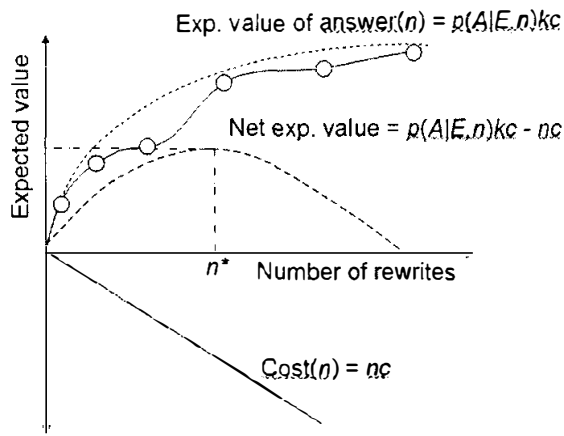

Figure 4: Key relationships in an idealized cost-benefit model for the case of decreasing returns in expected value of an answer with additional queries. The reality of non-monotonicity in expected value is highlighted with an irregular curve.

## 7 Empirical Study of Decision Making

We performed a set of experiments with AskMSR-DT, employing the utility model described above, to drive dynamic decisions about the best number of query rewrites to select. Given a query, AskMSR-DT generates all rewrites that would have been submitted in the legacy AskMSR system. The query rewrites are first ranked by the single-query models. Then, the ensemble of Bayesian models for different numbers of rewrites are employed in conjunction with the utility model to select the best number of rewrites to issue to a search engine. The search results are then passed to the answer composition stage of the system. The available actions are defined by the end-to-end performance models which were trained for thresholds of 1-10, 12, 15, and 20 rewrites.

Figure 5 shows the cost-benefit analysis graphically for the example query "Where is the Orinoco River?," with a cost per query of 1 and a correct answer valued at 10. In this case, the best decision available is to choose to submit 5 query rewrites.

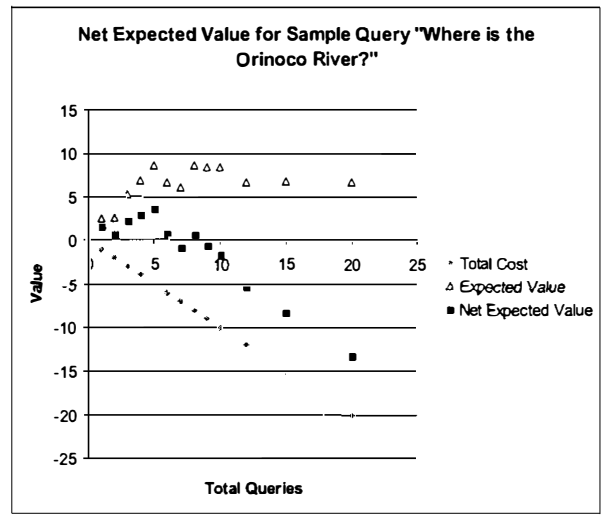

Figure 5: Expected value, cost, and net expected value of submitting different numbers of query rewrites for the question, "Where is the Orinoco River?"

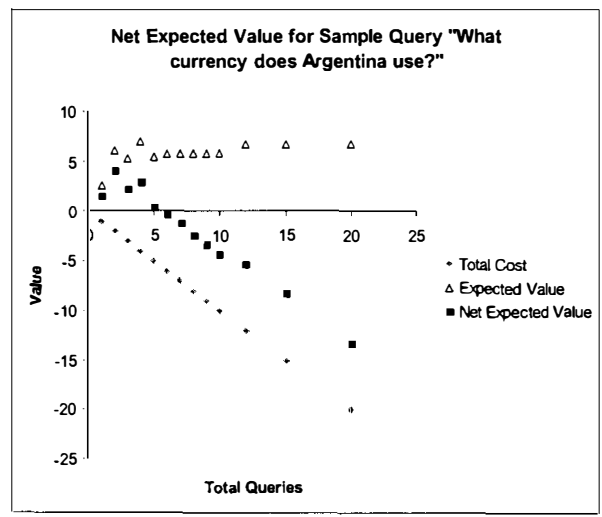

Figure 6: Expected value, cost, and net expected value of submitting different numbers of query rewrites for the question, "What currency does Argentina use?"

Figure 6 displays cost-benefit analysis for the query, "What currency does Argentina use?" for the same preference settings. With this policy, it is best to send 2 query rewrites to the search engine.

Table 4 shows the performance of the system over different baseline policies. In these fixed-cost runs, AskMSR is given a ceiling on the number of query rewrites it can use. In the first set of experiments, the system chooses randomly from the rewrites available for each query up to a threshold ($N$). In a second set of experiments, AskMSR was executed with a static policy of selecting $N$ rewrites from a list of query rewrites, ranked by the probabilistic query-quality score described in Section 5. A ceiling of 20 rewrites is roughly equal to the policy in the legacy AskMSR system, which had no limitation on rewrites, as only a few queries yield more than 20 rewrites. As highlighted in the table of results, sequencing queries by the query-quality score dominates the randomly ordered queries, demonstrating the value of using the query-quality score.

Table 4: Cost and accuracy for a set of baseline policies with fixed cost.

| Max Rewrites Per Question ($N$) | Total Cost | Correct Answers, Random Order | Correct Answers, Likelihood Order |
|---|---|---|---|
| $N = 1$ | 499 | 156 | 225 |
| $N = 2$ | 946 | 217 | 238 |
| $N = 3$ | 1383 | 243 | 254 |
| $N = 4$ | 1805 | 252 | 278 |
| $N = 5$ | 2186 | 272 | 282 |
| $N = 6$ | 2490 | 268 | 282 |
| $N = 7$ | 2738 | 272 | 282 |
| $N = 8$ | 2951 | 279 | 282 |
| $N = 9$ | 3103 | 276 | 282 |
| $N = 10$ | 3215 | 281 | 282 |
| $N = 12$ | 3334 | 281 | 283 |
| $N = 15$ | 3410 | 282 | 283 |
| $N = 20$ | 3426 | 283 | 283 |

We also used the ranked query rewrites for cost-benefit analysis. Table 5 compares the policy chosen by the cost-benefit analysis with two fixed policies, one using only conjunctional rewrites (top row) and the other using all rewrites (bottom row). Our results show good performance for the system using the cost-benefit control (middle row). With the cost-benefit analysis, the system answers nearly as many correct as the original, unbounded system (277 versus 283), while posing less than a third of the total queries used without control.

As a baseline comparison, the system was also executed with a fixed policy of using only the conjunctional rewrite for each question (first row, Table 5). This is useful because the conjunctional rewrite is the query reformulation that nearly always leads to the most results from the search-engine backend. This makes the conjunctional rewrite extremely valuable, as a greater set of intermediate results means a better chance of finding an answer. Our experiment shows that the conjunctional-query-only policy does fairly well, leading to 49% accuracy using only 499 total queries. However, this static policy is outperformed by the utility-directed system by a significant margin in terms of accuracy. Using the decision model, we achieve a 12% increase in correct answers at a cost of 680 additional queries. Another baseline is considers the current AskMSR system which submits all rewrites

Table 5: Cost and accuracy for AskMSR-DT versus the static policies of choosing only the conjunctional rewrite and using all rewrites.

| Rewrite Policy | Cost | Correct Answers (out of 499) |
|---|---|---|
| Conjunctional rewrites only | 499 | 247 |
| Cost-benefit $k=10, c=1$ | 1179 | 277 |
| All rewrites | 3426 | 283 |

Table 6 shows the cost-benefit relationships for four different values of $k$. With a value of $k=15$, we reach the performance of the current AskMSR system but with many fewer queries (1346 vs. 3426).

Table 6: Cost (total queries) and accuracy with cost-benefit decisions for four different settings of $k$, representing the value of a correct answer.

| Value of answer ($k$) | Cost | Correct answers |
|---|---|---|
| 5 | 603 | 253 |
| 10 | 1179 | 277 |
| 15 | 1346 | 283 |
| 20 | 1405 | 283 |

## 8 Summary and Future Work

We described our efforts to characterize and control a legacy Web-centric question-answering system. The methods demonstrate the promise of employing a layer of probabilistic analysis to guide the extraction of information from the Web in a Web-centric question answering system.

We employed two phases of machine learning to build Bayesian models that predict the likelihood of generating an accurate answer to questions, and showed how we can couple such predictive models with considerations of the value and costs of different web querying actions. The project demonstrates broadly the use of Bayesian procedures to understand and control the behavior of a heuristic system. More specifically, we demonstrated the use of probability and utility in guiding costly search actions undertaken by a Web-based question-answering system.

In ongoing work, we are studying how we can employ probabilistic analysis in several different ways to enhance additional components of the architecture and processes of question answering systems.

We are interested in extending the decision making considerations to consider issues of mixed-initiative interaction (Horvitz, 1999), where the decision models consider real-time input from users to refine or reformulate questions. Beyond selecting the best web-querying actions to take, we can include in cost-benefit analyses a consideration of when it would be best to ask a user to reformulate a question rather than expending effort on handling a query that would be expensive or likely to yield inaccurate results. In such an analysis, we consider an assessment of the cost of delay and effort associated with a reformulation and the likelihood that a reformulation would lead to a better result.

We seek to boost the predictive power of the models of answer accuracy by considering additional features of questions and query rewrites, and extending inference methods to acquire or reason about notions of topic, informational goals, and overall context of a user associated with a question. In relevant recent work, researchers learned models for predicting topic and high-level intentions associated with questions from tagged libraries of questions, posed by users of the Encarta online encyclopedia (Zukerman and Horvitz, 2001). The models provide predictions of the high-level information goals, topic, and desired level of detail of users, based on parts of speech and logical forms provided by an NLP parse of questions. There is opportunity to enhance the predictive power of models of the accuracy of answers to queries by incorporating the methods used in the prior work.

Beyond extending the probabilistic models of accuracy and expected value analysis, we are interested in refining the base question-answering system in several ways. Refinements of the base system could provide more power and control opportunities to the cost-benefit machinery. Promising refinements to the base system include introducing new variants of query rewrites and modifying methods for combining search results into candidate answers. In addition to guiding real-time question-answering procedures, the decision-analytic evaluative and control machinery can serve as a tool, enabling us to probe in an explicit manner the utility of making specific modifications to the base system.

Overall, we believe that pushing decision-theoretic analyses deeper into the operation of question answering systems will be fruitful in the development of methods for taking advantage of the massive, but unstructured knowledge of the Web. Beyond question-answering systems, we suspect that similar methods for introducing a "normalizing layer" of probabilistic analysis of accuracy coupled with utility-guided query control may be valuable for guiding the extraction of information from the Web and other large unstructured corpora, in support of a variety of tasks requiring information synthesis from large, unstructured corpora.

**Acknowledgments**

We thank David Weise for providing assistance with the Microsoft NLG statistical parser.